\documentclass[preprint]{vgtc} 
\ifpdf
  \pdfoutput=1\relax                   
  \usepackage{graphicx}                
  \DeclareGraphicsExtensions{.pdf,.png,.jpg,.jpeg} 
\else
  \usepackage{graphicx}                
  \DeclareGraphicsExtensions{.eps}     
\fi%

\graphicspath{{figures/}{pictures/}{images/}{./}}

\usepackage{microtype}                 
\PassOptionsToPackage{warn}{textcomp}  
\usepackage{textcomp}                  
\usepackage{mathptmx}                  
\usepackage{times}                     
\usepackage{cite}                      
\usepackage{tabu}                      
\usepackage{booktabs}                  

\onlineid{1355}
\vgtccategory{Workshop Paper}
\vgtcinsertpkg

\title{Teaching K-12 Classrooms Data Programming: A Three-Week Workshop with Online and Unplugged Activities}

\author{Alpay Sabuncuoğlu\thanks{e-mail: asabuncuoglu13@ku.edu.tr}\\
\scriptsize Koç University - Is Bank AI Center%
\and A. Evren Yantaç \thanks{e-mail: eyantac@ku.edu.tr}\\
\scriptsize Koç University - KUAR - Arçelik Research Center \\
\scriptsize KARMA Mixed Reality Lab
\and T. Metin Sezgin\thanks{e-mail: mtsezgin@ku.edu.tr}\\
\scriptsize Koç University - Is Bank AI Center
}

\abstract{This paper shares our experience in a three-session online workshop using a new web-based data programming environment, Marti. The programming environment uses a card-based programming strategy in both unplugged and online activities. Educators can use the physical cards in a board-game style or use the programming environment's mobile application to scan these cards and render the final visualization on their phones/tablets. The web environment also uses visual draggable cards for programming that can manipulate and visualize data. We used Marti and its offered unplugged activities in three sessions with 12 middle school and 12 high school students, focusing on the data fundamentals, analysis, and visualization. We assert that integrating unplugged-style pseudo-code creation and supporting a similar experience using the available devices have considerable potential for delivering equal and affordable data programming education for all. 

Our activity resources are available at \url{https://github.com/karton-project/marti}} 

\CCScatlist{
  \CCScatTwelve{Applied computing}{Interactive learning environments}{};
  \CCScatTwelve{Human-centered computing}{Field Studies}{}
}

\nocopyrightspace

\begin{document}



\maketitle

\section{Introduction}
Programming, data literacy, and analysis have become an interest in K-12 classrooms\cite{wolff-society}. Educators and policy-makers try to find the right tools to introduce data analysis and visualization methods to students. On the other hand, choosing the apt tools is difficult as the recent interest in data literacy increased the number of new applications. Most curricula traditionally suggest using spreadsheet applications (e.g., Microsoft Excel, LibreOffice Calc, Google Sheets) and prepare curricular activities accordingly \cite{tim2018computer}. There are also popular web-based educational tools like Tuvalabs \cite{tuvalabs}, CODAP \cite{codap}, and DataBasic \cite{databasic}. However, these methods demand several computers in the classroom, which is hard to attain in most underprivileged communities. In our workshops, we used a new web-based educational data programming tool that can support a range of unplugged and mobile activities, allowing us to involve children with limited or no access to computers or tablets.

\begin{figure}[h]
    \centering
    \includegraphics[width=\linewidth]{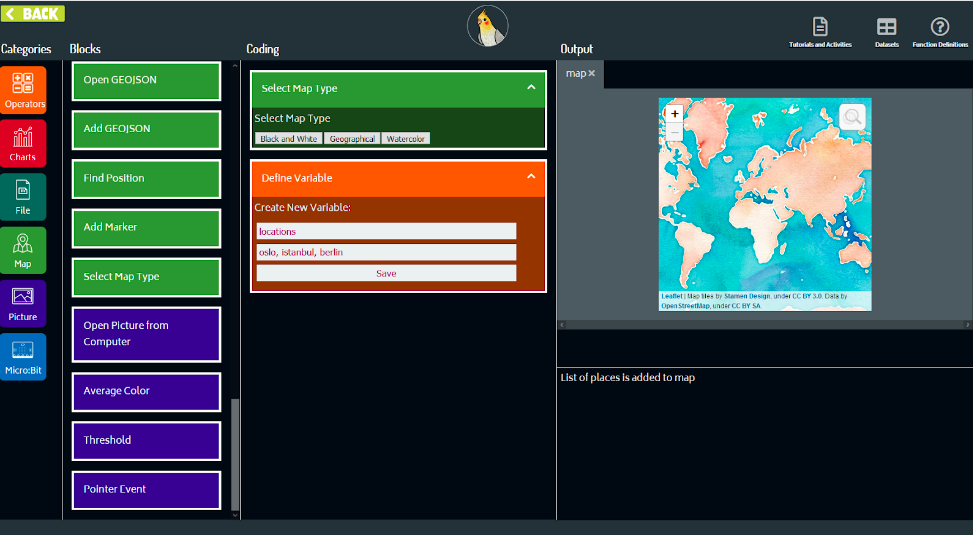}
    \caption{Web interface aims to align with the physical programming card design while showing a similar interaction with popular tools like Scratch \cite{scratch}. The screen has similar ordering and proportions with other drag-and-drop type web-based programming tools. }
    \label{fig:web}
\end{figure}

\begin{figure*}
  \centering
  \includegraphics[width=.8\linewidth]{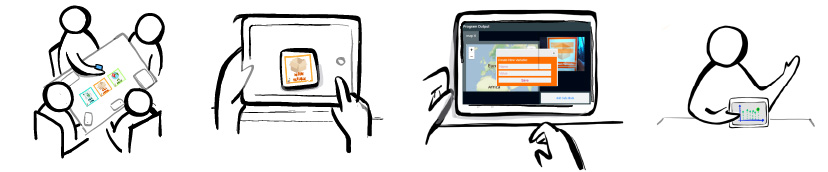}
  \caption{One possible workflow of completing a data programming activity using Marti. The programming environment involves programming cards and shared tablets. The mobile interface has two main areas: Output and Camera. Students can display the given physical activity cards to the device camera to add this programming block into their data flow. After editing the input, the corresponding change automatically occurs in the Output area.}
  \label{fig:teaser}
\end{figure*}

In our three-week online workshop program organized in collaboration with an NGO, we used Marti \cite{Sabuncuoglu2021Marti}, a new web-based data-programming environment. Marti's data programming environment contains (1) physical data-programming cards, (2) a companion mobile application, and (3) a web application with drag-and-drop visual programming blocks. This variety offers educators several usage scenarios; they can use these environments individually or combine the activities. For example, in an unplugged scenario, students can discuss the programming concepts and create pseudo-pipelines using only physical programming cards without needing a digital device. Supporting this kind of physical card environment enables working with students coming from low socioeconomic backgrounds. In another low-budget classroom scenario, students can use a shared smartphone to scan these cards and process data, as seen in Figure \ref{fig:teaser}. Marti mobile web application can recognize these cards using the device camera and opens a pop-up screen to enter the required input. Lastly, students can use Marti’s desktop web application. Figure \ref{fig:web} shows the web application which uses visual programming cards to build data manipulation pipelines and visualizations using its drag and drop interface. 

A total of 24 middle and high school students from various regions of Turkey attended to our three-session workshops. First, they used Marti’s unplugged activities using physical cards. Then, they built data pipelines and visualizations using the web interface. Overall, the students had different devices; some had only smartphones/tablets, while others had access to computers. Thus, Marti allowed us to complete various data-literacy activities regardless of the devices' screen sizes and capabilities. In this paper, we present the workshop activity flow, share our resources and report our experiences.

\section{Workshop Goals and Flow}
We organized a three-week program during the semester break of the students. In the design of the activity flow, we targeted three primary learning outcomes: \textbf{(1)} Increasing students’ awareness of the importance of data literacy in everyday life, \textbf{(2)} Making them fluent in choosing the proper analysis and visualization operations, \textbf{(3)} Showing a wide variety of data formats from diverse application areas.

We used both unplugged and web-application activities. We did not use the mobile application’s card recognition capabilities, as most students’ device camera were occupied by the Zoom application. Our activities are structured around Sustainable Development Goals (SDG)-related concepts to apply these data interpretation skills in real-life applications. Yet, we started the first day with Marti’s Football Player dataset to introduce more familiar names and concepts. Our content has board-game influences as evidence suggests that presenting the content in a structured and gamified environment boosts students’ motivation, comforts teachers, and accelerates the curriculum’s adaptation \cite{Barber2021}. The gamification of these activities also supports comprehending the new information in an engaging environment \cite{Barber2021}. In addition, the card-based programming structure allowed the easy integration of board game elements into the activities by design.

The following five questions drive the activity flow: 
\textbf{(1)} What is data?
\textbf{(2)} Where do we see data in real life?
\textbf{(3)} How can we use data to solve our problems?
\textbf{(4)} How can data visualization help to understand our problem better?
\textbf{(5)} What are the common fallacies in data interpretation?

\begin{figure}[hb]
    \centering
    \includegraphics[width=\linewidth]{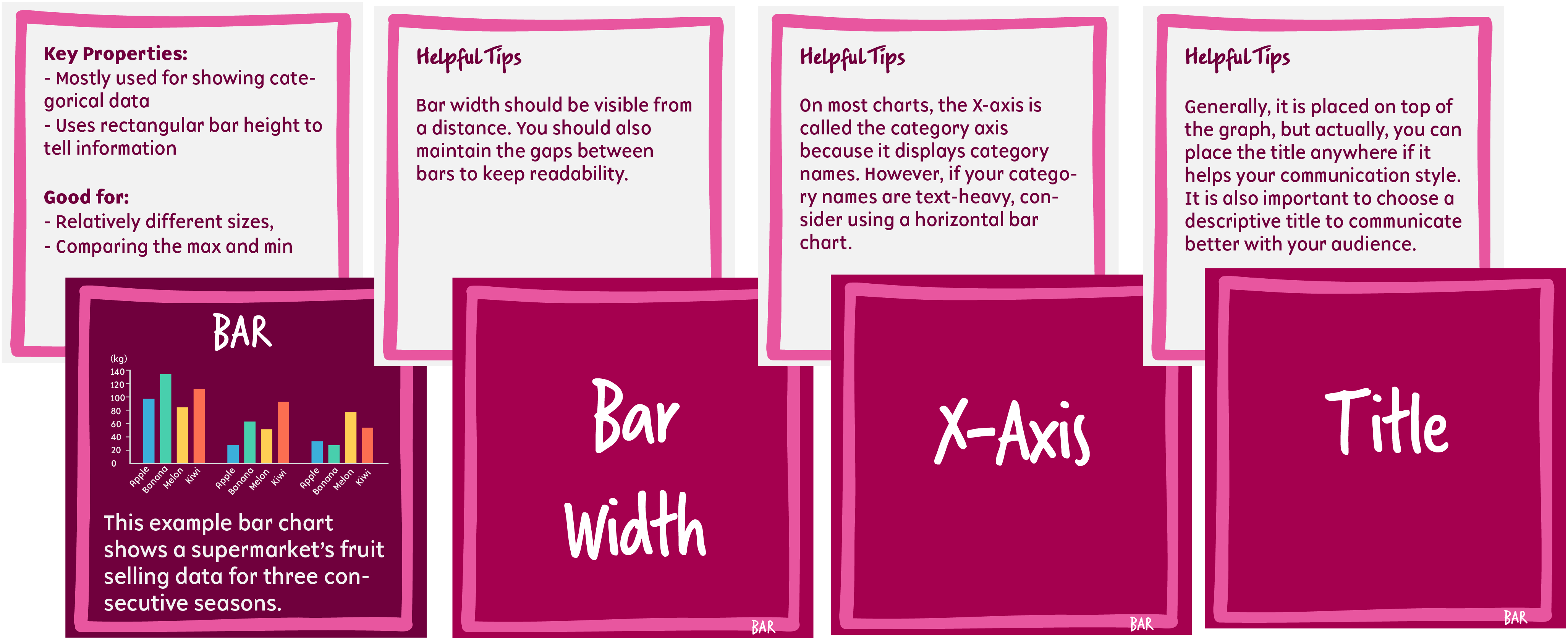}
    \caption{Each visualization card includes key helping information to support students in the chart selection and creation steps. This figure shows \textit{Bar Chart} card and three  \textit{Chart Element} cards. The chart card contains an example in the front and summarizes the key properties and tips. Each chart type has its element cards that summarize helpful tips to place the element on chart creation.}
    \label{fig:viscards}
\end{figure}

\begin{figure}[hb]
    \centering
    \includegraphics[width=\linewidth]{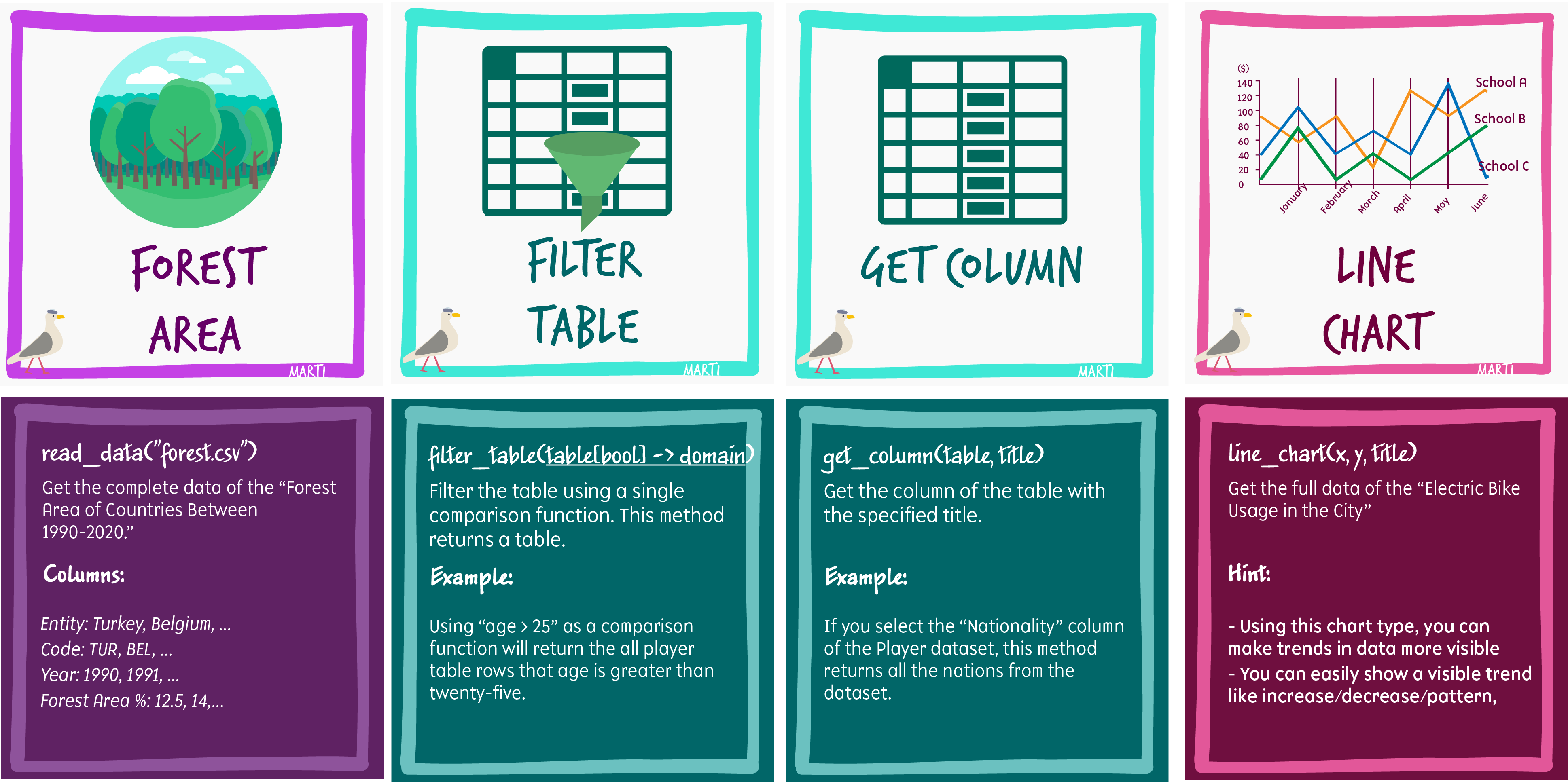}
    \caption{Students choose these four cards to create a high-level pipeline to generate a line chart that shows the change in forest areas of a country through years. Each category is colored separately to ease the choosing process. At the back of the cards, students can find the definition and an example usage. }
    \label{fig:card-program}
\end{figure}

Based on these starting questions, we narrated a gamified scenario about traveling around the world to solve various sustainability issues. We followed three types of tasks: (1) Creating pseudocodes that involve selecting and ordering the physical programming cards as seen in Figure \ref{fig:card-program}, (2) completing digital activities using the web application, and (3) visualization with unplugged cards and web interface. In these gamified tasks, the moderator can give points based on the solution’s correctness or considering the solution’s time.   In a classroom environment, groups also can compete with each other. Yet, due to the online setting’s limitations, we applied question-wise gamification in our workshops. For example, students were asked to create a chart with all the required elements like title and axis labels in the visualization part. When students could not complete the chart with the required components, they requested hints. Figure \ref{fig:viscards} shows three of the component cards that can be handed out when students asked for help. When they received the card, they lost a point but complete a missing ingredient of their chart.

\subsection{Method and Tools}
We conducted weekly meetings with 24 middle and high school students from a local non-governmental organization (NGO) for three weeks. Only six middle school students and eight high school students participated in all three sessions of our workshop; we used only their data in the analysis. Each session, we devoted two hours to use our programming cards, tools, and gamified curricular content. A moderator and three observers were present in each session. The observers took notes on two axes: Attention and understandability. Attention is described as learners’ active cognitive behavior on reception and response to stimuli \cite{Smith2009}. The observers tracked if students actively asked questions, followed the main flow, and showed general interest. We measured understandability based on students’ ease in using the activity content and comprehending the concepts. The independent observers were selected from senior-year students from a university’s education department with no relations with the authors.

Throughout the activities, we asked questions to students via \href{https://spiral.ac}{Spiral}, an online formative assessment tool, to make the Q\&A process more engaging. We did not conduct a traditional pre-tests and post-tests. Instead, we asked repetitive and similar questions through Spiral in the first and the last session. This kind of assessment strategy allowed us to use our pre and post-test questions as discussion starters, quickly gauge students’ knowledge levels and adapt the content.

Both middle and high school students had previous knowledge of simple statistics subjects such as drawing charts, applying basic math operations, or collecting data. These topics are 6th and 7th-grade subjects in their national math curriculum. However, our Q\&A results demonstrate that they did not remember most of these concepts.

\subsection{Introduction (Day 1)} 
We started by defining the fundamental terms such as algorithm, coding, data, and dataset. We first asked the students to guess the meanings of these terms via Spiral, then introduced the concepts one by one. Then, we shared daily uses of data such as social media applications, auto-correct keyboards, and video suggestions. We also showed some data stories from \textit{pudding.cool}, where journalists share interesting data stories. In this part, educators can introduce different stories by considering the group dynamics. In our workshop, we shared three interesting uses of data analysis and visualization: (1) A publishing editor’s use of book cover picture data to inform the next cover decision, \footnote{\url{https://pudding.cool/2019/07/book-covers/}} (2) A social scientist’s analysis of lifestyle and political events by examining people’s hair length data. \footnote{\url{https://pudding.cool/2019/11/big-hair/}} (3) Web’s changing laugh styles in the last decade. \footnote{\url{https://pudding.cool/2019/10/laugh/}}

We finally introduced the Football Player dataset, which we prepared as a subset of EA Sports FIFA 2018 Game’s Players that includes players’ name, country, age, potential, and overall point attributes. After exploring the dataset attributes, we introduced basic operations like filtering data, selecting a column, saving a variable. Then, we tested each operation one by one using the dataset and showed how each command requires different inputs and results in different outputs.

\subsection{Data Collection and Datasets (Day 1 and 2)} 
A popular data collection activity is asking students to conduct surveys with their friends. In the online studies, we did not conduct this activity due to logistic complexity. Yet, this kind of familiar activities results in meaningful learning as the content is presented within a personally relevant setting \cite{Chu20191}. We did not use Zoom polling as it was time-consuming, and we did not want to spend too much time on complex and distractive operations. Instead, as a familiar dataset, we introduced Football Player dataset as explained above. 

On Day 2 and 3, we continued with Marti's offered Sustainable Development Goals (SDG) activities and the simplified \textit{plastic consumption} and \textit{forest area} data. In our dataset selection, we focused on two key elements. First, the concept should be familiar but exciting for the students. Second, the data should include different primitive types such as strings and integers to use all our data programming environment’s functionality. In these terms, we selected datasets that contain exciting facts related to science topics in their curriculum and touches on SDG-related issues.  

\subsection{Data Analysis and Visualization (Day 1, 2, and 3)} 
This set of activities starts with introducing the UN's Sustainable Development Goals (SDG). Students first analyze the given data, then choose the appropriate visualization method to communicate their results. Aligned with recent literature, we focused on teaching visualization as a rethinking of numbers and ideas to communicate the stories \cite{wolfe2015}. Our SDG-related activities include datasets for \textit{Forest Area}, \textit{City Bikes}, \textit{Research Budgets}, and \textit{Plastic Production}. These datasets link science and SDG concepts that are also part of the middle and high school curriculum. Each dataset is determined to fulfill curricular needs in our gamified education flow. We prepared eight data analysis questions and two visualization questions aligned with each dataset.

While determining the visualization questions, we focused on choosing the suitable visualization method that could communicate our ideas well. Marti programming environment supports a limited set of visualization operations (table, line, bar, pie, and map charts), so we were asked to visualize the results with these supported operations. We used Marti's visualization operation cards (Figure \ref{fig:viscards}) to remind students of the elements of drawing a chart.

\subsection{Data Fallacies (Day 3)} 
We introduced data fallacies at the end of the workshop, as it has become more and more important for everyday media users. First, we discussed what sounds wrong in their visualization and representation through the analysis activities while presenting the results. Next, we introduced some example cases using Marti's data fallacies cards. Marti introduces nine data fallacy examples and three samples for each card. These fallacies are cherry-picking, survivorship bias, false causality, gerrymandering, sampling bias, overfitting, gambler's fallacy, Hawthorne effect, and danger of summary metrics. However, we only introduced cherry-picking and the danger of summary metrics, as the time is limited. Finally, we discussed if one data fallacy occurred during our workshop or if they encountered any examples. 

\subsection{Activity Questions}
We asked for definitions of basic terms, creating some pseudocode to solve data-driven problems and draw charts. We shared all questions for each day with the following question type abbreviations,

\begin{itemize}
    \setlength\itemsep{-0.3em}
    \item OE: Spiral - Open Ended
    \item MC-N: Spiral - Multiple Choice with N options
    \item CNV: Spiral - Canvas
    \item M: Solved in Marti, Pasted the Result to Spiral
\end{itemize}

\subsubsection{Day 1}
Day 1 questions aim to learn the demographics and can be interpreted as pre-test.
\begin{enumerate}
    \setlength\itemsep{-0.3em}
    \item Which grade are you in? (MC-2)
    \item Which device do you use to connect with Zoom? (MC-3)
    \item What is data? (OE)
    \item Draw a bar graph that shows the count of the same-age students. (CNV)
    \item Which cards should we choose to find the players from Argentina? (M)
    \item Which cards should we choose to find the players from Real Madrid? (M)
    \item How is your relationship with technology? (MC-4)
    \item How do you find the applications that we used today? (MC-4)
\end{enumerate}

\subsubsection{Day 2}
These questions aim to keep the discussion active, and understand current knowledge.
\begin{enumerate}
    \setlength\itemsep{-0.3em}
    \item How many goals do the UN present to achieve sustainable development? (MC-4)
    \item Draw the table as a pie chart. (CNV)
    \item What is the comparison function to get the table that only contains Brazil (M)
    \item Draw a graph that shows the change of the area from 1990 to 2015 in Brazil. (CNV)
    \item Can you do all the things that come to your mind while using the application? (MC-4)
\end{enumerate}

\subsubsection{Day 3}
Day 3 questions aim to understand students' development and can be interpreted as post-test.
\begin{enumerate}
    \setlength\itemsep{-0.3em}
    \item What is data? (OE)
    \item Which cards can we use to chart the forest areas of Brazil from 1990 to 2015? (M)
    \item Draw the change in forest area of Brazil between 1990 and 2015.
    \item Which programming cards should we use to find FIFA 2018 players over the age of 29? (M)
    \item Which programming cards should we use to find FIFA 2018 players over the potential of 90? (M)
    \item Which programming cards should we use to find the average age of Spanish players? (M)
    \item Which programming cards should we use to find the oldest player? (M)
    \item Which programming cards should we use to find the least potential? (M)

\end{enumerate}

\subsection{Limitations}
The programming cards and activity flow was initially intended to be used in the physical classroom experience. However, we observed that using the card metaphor and creating pseudo-codes with digital cards also engaged the students in online studies. 

In an online collaborative activity, student-educator and student-learning content engagement are challenging to achieve and maintain. Therefore, during the studies, we kept the pace slow. In this case, some students got bored, and some others struggled with technical difficulties or lack of experience. Further, we did not force them to open their microphones or camera, so our observations are only limited to their reactions when their camera and microphone was open.

\section{Analysis and Discussion}
In the analysis step, we categorized the observer notes based on attention and understandability axis. Each observer assigned an attention and understandability score (1-10) in the end of each session and take regular notes at every five minutes. Overall, the attention score has \(\mu\) = 8.4 and \(\sigma\) = 1.7 and understandability score has \(\mu\) = 8.3 and \(\sigma\) = 1.3. Although this scores are promising and motivates us to conduct our research in our current direction, the number of students and observers limits interpreting the quantitative results. 

\subsection{Observer Notes}
Observers asserted that using our activity flow increased students’ motivation and helped them grasp data literacy. Giving unique and fun examples, summarizing the content, building communication with each student, gradually increasing the difficulty by integrating card-based pseudocode creation are all listed as best practices. We summarized the observer notes in four categories: Growing attention, losing attention, making concepts tangible, and making concepts confusing.

\begin{figure*}[ht]
    \centering
    \includegraphics[width=\textwidth]{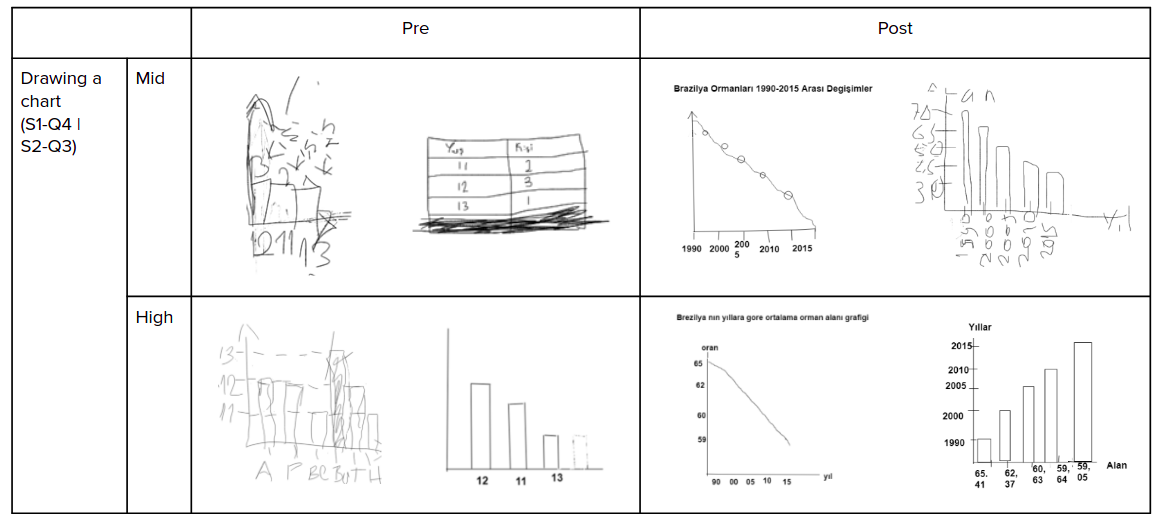}
    \caption{Student's chart drawings in the first and last meeting. We shared the answers that is drawn on Spiral's online canvas. Other students showed their answers through camera.}
    \label{test-charts}
\end{figure*}

\noindent\textit{\textbf{Growing attention/excitement:} Using real-life examples, 'cool' new concepts, switching between cards and digital activities, design of programming cards}

\begin{itemize}
    \setlength\itemsep{0em}
    \item \textit{Using 'cool' real-life examples:} Describing the working principles behind familiar real-life data-driven mechanisms, such as explaining how YouTube works or how a real-life robot can find its way raised students’ attention. The shared data stories were 'cool' and 'familiar.’ These 'cool' concepts grew their attention in transition to SDG-related concepts. The SDG-related activities were such as rainforest area declination and matched their interest as it was a common issue for everyone.
    
    \item \textit{Switching between cards and digital activities:} Using different yet familiar learning modalities (i.e., cards, canvas, quizzes) helped them stay focused on the tasks. Students used cards and a web interface together, which helped them to stay more focused during the activities. We included gamification with small challenges, such as predicting the hue value of the given color. 
    
    \item \textit{Design of the programming cards:} The shape and colors of the data programming cards positively impacted students’ attention. Using cards like pseudocode elements helped students complete activities more effortless, which resulted in higher participation.
\end{itemize}

\noindent\textit{\textbf{Losing attention/excitement:} Switching between online tasks, Having mixed devices}

\begin{itemize}
    \setlength\itemsep{0em}
    \item \textit{Switching between online tasks:} In the first meeting, technical difficulties like opening the Zoom meeting, visiting links from Zoom chat, switching tabs between Zoom, Spiral, and web applications were challenging. Nevertheless, students become more fluent in switching between tasks after the first meeting.
    
    \item \textit{Selecting files on mobile:} In the first week, middle school phone users had a hard time downloading and using the dataset. Then, we started using "Open CSV with a web link" programming card and included our all datasets as online resources.
\end{itemize}

\noindent\textbf{\textit{Making concepts tangible/easier to understand:}} Giving examples from real-life, Asking basic what and how questions, Step by step execution

\begin{itemize}
    \setlength\itemsep{0em}
    \item \textit{Giving examples from real-life:} Defining algorithmic steps with an analogy of making a sandwich process and sharing real-life data stories from the pudding website helped students making concepts more tangible. In addition, hearing familiar football player names increased the attention of students in first day. All observers mentioned that integrating stories from this website raised attention significantly.
    
    \item \textit{Asking basic what and how questions:} Boosting the discussion via Spiral and gradually increasing the question difficulty helped students understand each step and raised their attention.
    
    \item \textit{Step by step execution:} Another advantage of using pseudocode data cards was teaching one concept at a time.
\end{itemize}

\noindent\textbf{\textit{Making concepts more confusing:}} Shallow explanation of details, Switching tasks ‘too fast’

\begin{itemize}
    \setlength\itemsep{0em}
    \item \textit{Shallow explanation of details:} Even if one step of the algorithm is explained shallowly, it could confuse students. For example, one of the activities involved the filter value update in getting a completely different subset.
    
    \item \textit{Switching tasks ‘too fast’:} Our aim was to growing some attention by just changing one value to obtain a different subset. However, students were confused. A collaborative/interactive Q\&A can help to explain these details. In particular, switching tasks in an online setting ‘too fast’, such as changing windows or tasks, can confuse students.

\end{itemize}

\subsection{Spiral Questions Analysis}
We asked basic definitions, card-ordering tasks, and drawing charts in Spiral activities. In the first workshop, we asked two card-ordering questions and two statistics questions. However, none of the middle school students could answer the basic definition and statistics questions, such as averaging a part of the data in the first session. Although it was students’ first time learning about these cards, they could predict the correct cards on Day 1. For example, half of the middle school students predicted the correct cards to solve question 5 of Day 1. High school students are more experienced in data-related operations, they adapted the system quickly, and 5/8 students could answer this question. We can interpret this observation as the success of card-based methodology in teaching. On Day 3, more than half (3-5/6) of middle school students answered questions 4,5, and 6 correctly, which shows a visible improvement over the pre-test scores. High school students did not show a significant improvement, but our observer notes revealed that they had become more confident throughout the process.

We also asked students to draw a chart at the beginning and end of the studies. We observed that the chart programming block helped them remember the fundamental steps (e.g., adding a title, giving label names to x and y axes) to draw a chart. Figure \ref{test-charts} shows the middle and high school students’ answers in the first and second meeting to a chart drawing question. Drawing a chart in an algorithmic style pushed them to repeat the same structure repeatedly, which naturally aligns with spaced repetition learning technique.

\section{Findings and Discussion}
We grouped our findings under five categories based on programming environment functionalities, group dynamics, and pedagogical appropriateness.

\begin{itemize}
    \setlength\itemsep{0em}
    \item \textit{Cards modularity:} Our observations revealed that executing the code command by command benefits both teachers and students. Students can quickly experiment with programming operations and understand the impact of each function in the digital pipeline before compiling the code. Educators can structure the activities and create a modular flow as each card is designed similarly. 
    
    \item \textit{Switching tasks and multimodality:} Using the cards in both physical and digital forms helped students to follow the discussed subjects. In our studies, all students answered the questions and participated in the discussion that involved programming cards. Students quickly adopted using the cards to represent the solution abstractly. The simple use of the cards enabled students to focus on the solution rather than technical details. Thinking about the main steps in the first place initiated a fruitful discussion where we discussed each step in detail. We continue our research using card methodology by including this interaction in a board game or book structure.
    
    \item \textit{Choosing the right dataset:} The students quickly understood the flow of our programming environment and could easily use it in the following workshops. Both middle and high school students quickly learned to utilize our datasets, programming cards, and interface elements. The observer notes show that the understandability increases by integrating real-life data stories and supporting gradual difficulty increase. Interdisciplinary content in current curricula includes complex elements that require learning technical details to analyze the data, weakening teachers’ confidence. Our environment promoted data programming in an easy-to-grasp fashion.
    
    \item \textit{Pedagogical appropriateness:} Students could quickly understand the cards’ functions and the system’s capabilities in the workshops, demonstrating that the application design is appropriate for middle and high school groups. The design elements of cards such as colors, icons, and abstraction level of input fields helped students remember the card’s function. In addition, the abstractions of the high-level data operations were appropriate as they were semantically similar to current popular data science languages and easy to understand at the same time.
    
    \item \textit{Improving gamification:} Current data programming cards and companion activity cards support the gamification to some extent. We released all the activity cards on \href{https://github.com/karton-project/marti}{Github} and would like to see different uses of these cards.
\end{itemize}

\section{Conclusion}

In this paper, we presented our workshops with 12 middle school and 12 high school students, focusing on the data fundamentals, analysis, and visualization. We used Marti and its offered unplugged activities. Throughout the workshops, our observations revealed that integrating unplugged-style pseudo-code creation and supporting a similar experience using the available devices have considerable potential for delivering equal and affordable data programming education for all. Although we could not use all interface elements due to limitations of the online setting, we demonstrate that using a card-based programming system engages students in the activity by naturally supporting step-by-step execution. We believe that our online resources will accelerate the integration of card-based tools in both classroom and hybrid activities.

\acknowledgments{The authors gratefully acknowledge that this work was supported by TUBITAK [Grant Number 218K436] and Koç University-İs Bank AI Center. We would also like to thank all the educators, volunteers and students from TOÇEV.}

\bibliographystyle{abbrv-doi}

\bibliography{template}
\end{document}